\documentclass[runningheads]{llncs}
\usepackage[T1]{fontenc}
\usepackage{graphicx}
\newcommand\blfootnote[1]{%
  \begingroup
  \renewcommand\thefootnote{}\footnote{#1}%
  \addtocounter{footnote}{-1}%
  \endgroup
}
\usepackage{marvosym}
\usepackage{booktabs}
\usepackage{pgfplots}
\usepackage{graphicx}
\usepackage{url}

\usepackage{breakurl}
\usepackage[breaklinks]{hyperref}
\usepackage{amsmath}
\usepackage{float}
\usepackage{pifont}
\newcommand{\cmark}{\ding{51}}
\newcommand{\xmark}{\ding{55}}

\usepackage[acronym]{glossaries}

\newacronym{ann}{ANN}{Artificial Neural Network}
\newacronym{ar}{AR}{Augmented Reality}
\newacronym{bnn}{BNN}{Binarized Neural Networks}
\newacronym{bwn}{BWN}{Binary Weighted Network}
\newacronym{bram}{BRAM}{Block Random Access Memory}
\newacronym{cpu}{CPU}{Central Processing Unit}
\newacronym{cnn}{CNN}{Convolutional Neural Network}
\newacronym{dma}{DMA}{Direct Memory Access}
\newacronym{dram}{DRAM}{Dynamic Random Access Memory}
\newacronym{dnn}{DNN}{Deep Neural Network}
\newacronym{dl}{DL}{Deep Learning}
\newacronym{elm}{ELM}{Extreme Learning Machine}
\newacronym{fpga}{FPGA}{Field-Programmable Gate Array}
\newacronym{fpu}{FPU}{Floating Point Unit}
\newacronym{fp}{FP}{Floating-Point}
\newacronym{fps}{FPS}{Farthest Point Sampling}
\newacronym{fel}{FEL}{Federated Edge Learning}
\newacronym{gpu}{GPU}{Graphics Processing Unit}
\newacronym{hmi}{HMI}{Human Machine Interface}
\newacronym{hmm}{HMM}{Hidden Markov Model}
\newacronym{hls}{HLS}{High Level Synthesis}
\newacronym{imu}{IMU}{Inertial Measurement Unit}
\newacronym{knn}{KNN}{K-Nearest Neighbor}
\newacronym{lfsr}{LFSR}{Linear Feedback Shift Register}
\newacronym{lidar}{LiDAR}{Light Detection and Ranging}
\newacronym{mlp}{MLP}{Multi-Layer Perceptron}
\newacronym{mac}{MAC}{Multiply-Accumulate}
\newacronym{ml}{ML}{Machine Learning}
\newacronym{ma}{mA}{Mean Accuracy}
\newacronym{npu}{NPU}{Neural Processing Unit}
\newacronym{nn}{NN}{Neural Network}
\newacronym{os}{OS}{Operating System}
\newacronym{oa}{OA}{Overall Accuracy}
\newacronym{pl}{PL}{Programmable Logic}
\newacronym{ps}{PS}{Processing System}
\newacronym{pe}{PE}{Processing Element}
\newacronym{ptq}{PTQ}{Post-Training Quantization}
\newacronym{qat}{QAT}{Quantization-Aware Training}
\newacronym{roi}{ROI}{Regions of Interest}
\newacronym{rts}{RTS}{repeated tree search}
\newacronym{sram}{SRAM}{Static Random Access Memory}
\newacronym{sps}{SPS}{Samples per Second}
\newacronym{svm}{SVM}{Support Vector Machine}
\newacronym{simd}{SIMD}{Single Instruction Multiple Data}
\newacronym{tpu}{TPU}{Tensor Processing Unit}
\newacronym{tnn}{TNN}{Ternary Neural Network}
\newacronym{urs}{URS}{Uniform Random Sampling}
\newacronym{vr}{VR}{Virtual Reality}

\begin{document}
\title{HLS4PC: A Parametrizable Framework For Accelerating Point-Based 3D Point Cloud Models on FPGA}
\titlerunning{Accelerating Point-Based 3D Point Cloud Models on FPGA}
%
\author{Amur Saqib Pal\inst{1}$^*$ \and
Muhammad Mohsin Ghaffar \inst{2}$^*$\textsuperscript{(\Letter)} \and
Faisal Shafait\inst{1} \and
Christian~Weis\inst{2} \and
Norbert Wehn\inst{2}}
\authorrunning{A. S. Pal \& M. M. Ghaffar et al.}
%
\institute{National University of Sciences and Technology, 44000 Islamabad, Pakistan 
\email{\{apal.bee19seecs,faisal.shafait\}@seecs.edu.pk}\and
Microelectronic Systems Design Research Group, RPTU Kaiserslautern-Landau, 67663 Kaiserslautern, Germany
\email{\{mohsin.ghaffar,christian.weis,norbert.wehn\}@rptu.de}}
\maketitle              
\def\thefootnote{*}\footnotetext{These authors contributed equally to this work}
\begin{abstract}

Point-based 3D point cloud models employ computation and memory intensive mapping functions alongside \gls{nn} layers for classification/segmentation, and are executed on server-grade \glspl{gpu}. The sparse, and unstructured nature of 3D point cloud data leads to high memory and computational demand, hindering real-time performance in safety-critical applications due to \gls{gpu} under-utilization. To address this challenge, we present HLS4PC, a parameterizable \gls{hls} framework for \gls{fpga} acceleration. Our approach leverages \gls{fpga} parallelization and algorithmic optimizations to enable efficient fixed-point implementations of both mapping and \gls{nn} functions.
We explore several hardware-aware compression techniques on a state-of-the-art PointMLP-Elite model, including replacing \gls{fps} with \gls{urs}, parameter quantization, layer fusion, and input-points pruning, yielding PointMLP-Lite, a $4\times$ less complex variant with only $\sim2\%$ accuracy drop on ModelNet40. Secondly, we demonstrate that the \gls{fpga} acceleration of the PointMLP-Lite results in $3.56\times$ higher throughput than previous works. Furthermore, our implementation achieves $2.3\times$ and $22\times$ higher throughput compared to the \gls{gpu} and \gls{cpu} implementations, respectively. The code of the HLS4PC framework will be available at: \href{https://github.com/dll-ncai/HLS4PC}{https://github.com/dll-ncai/HLS4PC}. \blfootnote{The research reported in this work is partially supported by the Carl Zeiss Stiftung, Germany, under the Sustainable Embedded AI project (P2021-02-009).}
 
\end{abstract}

\section{Introduction}

\glsresetall

In recent years, 3D point cloud data from \gls{lidar} or RGB-D sensors is increasingly used in applications such as autonomous driving, robotics, drones, 3D reconstruction, \gls{vr}/\gls{ar} head-sets and even the iPhone 16 Pro. Processing 3D point clouds is challenging due to their sparsity, with unevenly distributed data points in 3D space. Since classification and segmentation are vital for safety-critical and real-time applications, these models must meet strict throughput demands. For instance, the throughput requirement for an end-to-end level-5 autonomous driving solution is estimated to be at 2,000 TOPS~\cite{nvidia_thor}.

In the literature, researchers have proposed projection-based, volumetric-based, mesh-based, and point-based methods for classification and segmentation of 3D point cloud data~\cite{Sarker_2024}. The point-based approaches~\cite{POINTNET,POINTMLP,DUALNET} dominate due to the ability to operate directly on raw 3D data, achieving up to 5\% higher accuracy and lower complexity~\cite{Lin_2021}. These models combine \gls{dnn} layers with mapping functions like \gls{fps} and \gls{knn} to extract features from unordered, sparse 3D data. While \glspl{gpu} excel at dense matrix operations, they struggle with irregular mapping functions due to data sparsity~\cite{POINTNET}, leading to resource under-utilization. A limited number of prior studies have explored the use of Application-Specific Accelerators~\cite{Lin_2021} to enhance the throughput of 3D point cloud models. Although, these accelerators can deliver high throughput, they inherently lack flexibility, making it challenging to deploy evolving and mixed-precision models.  

In contrast, \gls{fpga} allow precision to be configured at compile-time, support mixed-precision acceleration without hardware redesign. Additionally, new layers or functions can be added by updating hardware libraries or reconfiguring logic blocks. These advantages make \glspl{fpga} ideal for 3D point cloud processing, where models evolve rapidly. While various \gls{fpga}-based DNN frameworks exist~\cite{HLS4ML,FPGACONVNET}, they cannot accelerate 3D point cloud models due to their lack of support for point cloud mapping functions. To bridge this gap, we propose a parameterizable mixed-precision dataflow-based streaming framework for acceleration of 3D point cloud models on \gls{fpga} presented as HLS4PC. We perform an in-depth investigation of the effects of compression techniques such as input point pruning, quantization, and layer fusion combined with hardware-aware mapping functions on model accuracy, utilizing the ModelNet40 and ScanObjectNN as benchmarks. Based on these explorations, we introduce PointMLP-Lite, a $4\times$ smaller version of PointMLP-Elite, with only a $\sim2\%$ accuracy drop. We deploy PointMLP-Lite on a ZC706 development board, achieving $3.56\times$ higher throughput than prior work, and outperforming \gls{gpu} and \gls{cpu} in terms of throughput by $2.3\times$ and $22\times$ respectively.

\section{HLS4PC Workflow}
\label{hardware_implementation}

\begin{figure}[!ht]
\centering
\includegraphics[width=0.9\textwidth]{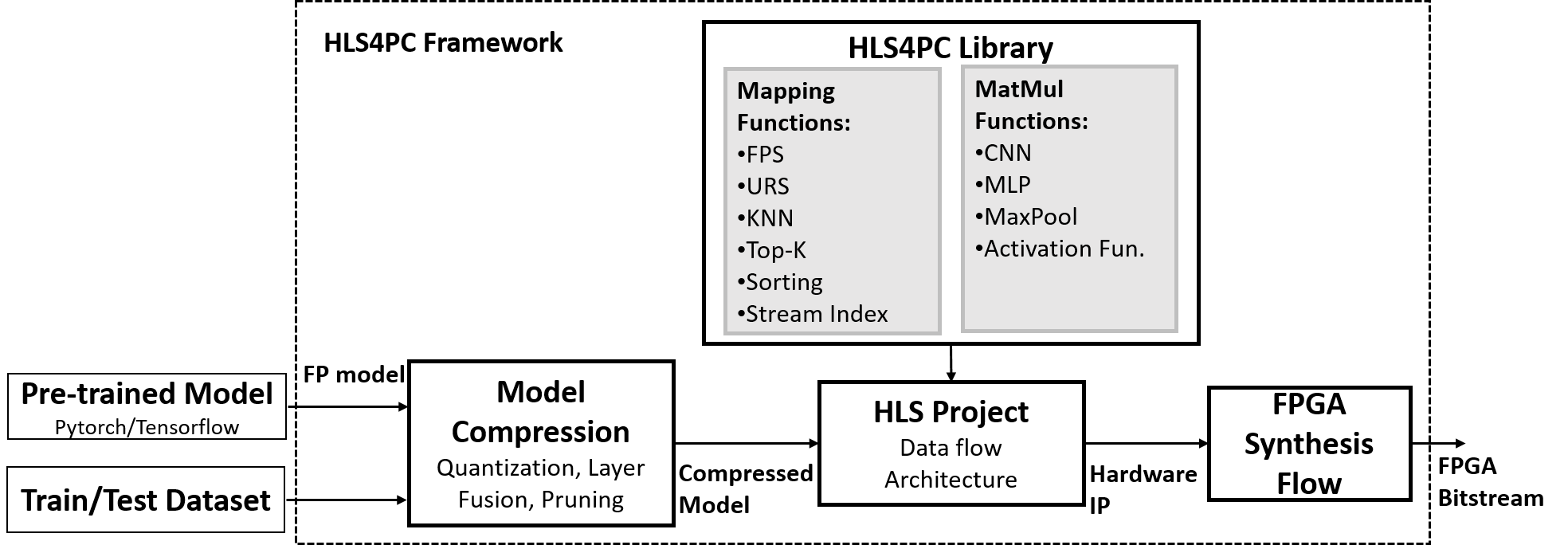}
\caption{The HLS4PC framework workflow.} \label{fig:pcl_fpga_block_diagram}
\end{figure}

In this section, we introduce HLS4PC, a framework designed to accelerate 3D point cloud models on \glspl{fpga}. HLS4PC leverages \gls{hls} to translate a high-level classification/segmentation 3D point cloud model into a hardware implementation. This approach allows for the rapid prototyping and deployment of customized hardware accelerators tailored to the specific requirements (precision, parallelism, and throughput). This framework consists of a fixed-point parameterizable HLS4PC library, which can be used to deploy 3D point cloud mapping (such as \gls{fps}, \gls{urs}, sorting and \gls{knn}) and MatMul (\gls{cnn}, \gls{mlp}, and max-pooling) functions, as illustrated in Fig. \ref{fig:pcl_fpga_block_diagram}. All these functions and layers are implemented using a streaming-based dataflow architecture approach. The HLS4PC framework accepts a pre-trained \gls{fp} model along with the dataset as input. It employs quantization-aware training to compress the model, uses the HLS4PC library to generate an \gls{hls} template, synthesizes the design, and ultimately produces an \gls{fpga} bitstream for deployment.    

\subsection{Mapping Functions}
Point-based methods~\cite{POINTMLP,POINTNETPLUS,POINTASNL} commonly use \gls{fps} to select \gls{roi} for local feature extraction, but its sequential nature and frequent distance updates make it compute and memory intensive. Previous works~\cite{POINTHOP,POINTCNN} have explored \gls{urs} for the accuracy/resolution trade-off and data augmentation. Unlike \gls{fps}, \gls{urs} selects points randomly, making it hardware-friendly. Rather than using \gls{urs} solely for augmentation, we replace \gls{fps} with \gls{urs} in the model architecture itself. Although \gls{urs} introduces stochasticity that hinders convergence, we mitigate this by increasing the batch size (up to 256) and training duration (up to 1,000 epochs), enabling the model to learn stable, geometry-invariant features. For hardware implementation, we implement \gls{urs} using a pseudo-random number generator based on \glspl{lfsr}. We seed our training experiments, initialize the \glspl{lfsr} with the same starting states, and use primitive polynomials to define their feedback mechanism.  

\begin{figure}[t]
  \centering
  \includegraphics[width=0.75\textwidth]{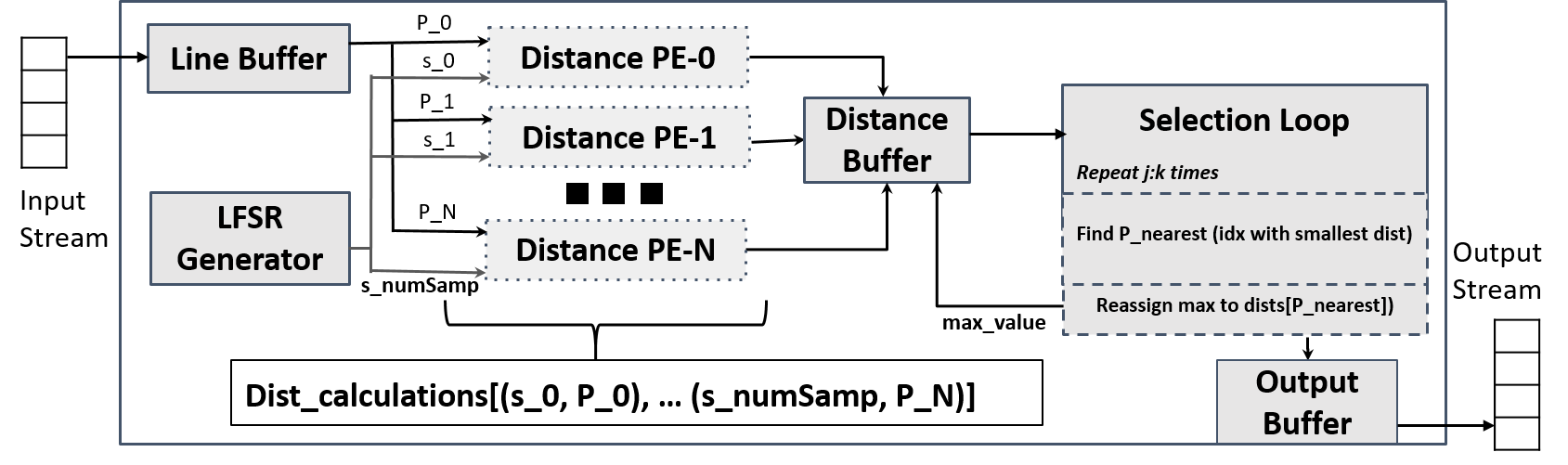}
  \caption{Architecture of the \gls{knn} algorithm}
  \label{fig:knn_algo}
\end{figure}

We take a multi-\gls{pe} approach to implementing \gls{knn} and leverage several \gls{hls} optimization techniques such as buffer partitioning, loop unrolling, and pipelining in our design. The implementation, shown in Fig.~\ref{fig:knn_algo}, uses a set of \(X\) parallelized distance calculation units referred as \textit{distance PE-0} to \textit{distance PE-N}, which take \(numSamp\) \gls{lfsr} samples (\textit{s\_0} to \textit{s\_numSamp}) and \(N\) points/features (\textit{P\_0} to \textit{P\_N}) as input. For every sample, the distance from each point in the input is calculated and stored in a \textit{distance buffer}. A selection sort-style module then finds \gls{knn} of each sample. The index of the point with the smallest distance to the sample is identified, and the distance value of that neighboring point is reassigned the maximum numeric limit of its fixed-point representation. This process is repeated \(k\) times for \(numSamp\) samples (the total number of samples), where \(k\) is the number of neighbors required for each sample. These neighbor sets collectively represent the isolated local region which is used for feature extraction further in the network. 
In our implementation, we use \(k=16\), \(X\)=4, \(numSamp\) $\in \{256, 128, 64, 32\}$ for 4 stages of the PointMLP-Lite topology.

\begin{figure}[t]
\centering
\includegraphics[width=0.75\textwidth]{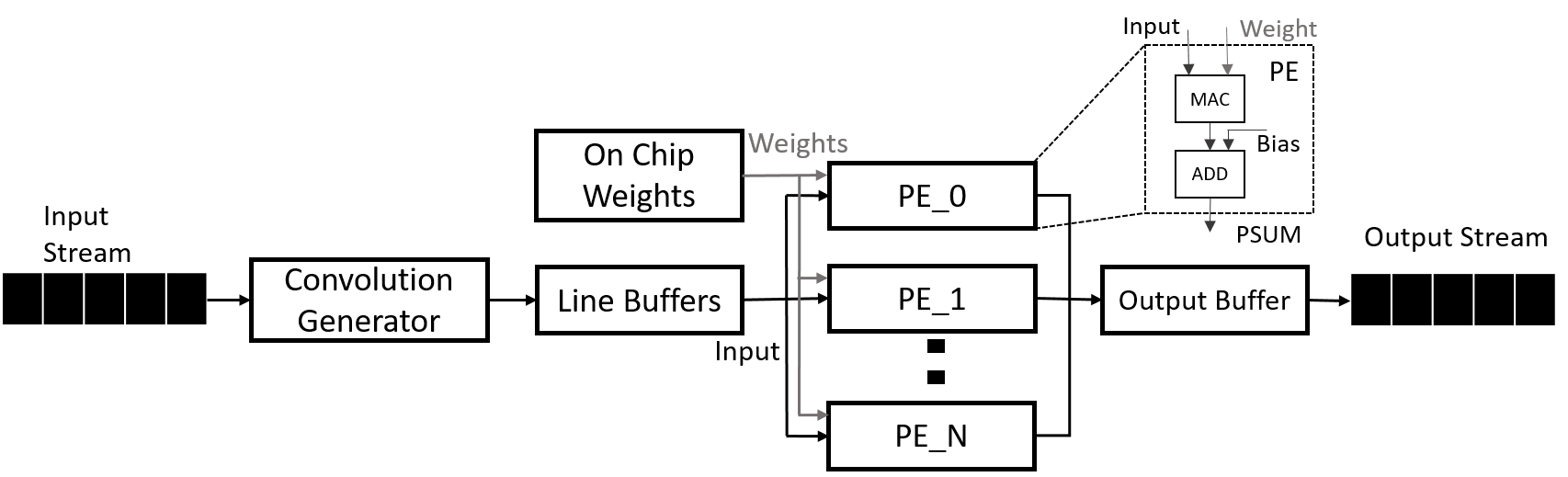}
\caption{Architecture of the convolution layer} \label{fig:convolution}
\end{figure}

\subsection{MatMul Functions}
\gls{cnn} and \gls{mlp} layers are implemented using a streaming dataflow architecture, where each layer is a distinct hardware module with configurable parallelism, as shown in Fig.~\ref{fig:convolution}. The number of \glspl{pe}, labeled $PE_0$ to $PE_N$, controls the degree of parallel \gls{mac} operations per layer. These compile-time parameters enable customization based on resource availability and throughput requirement. Since the most complex layer dictates overall throughput, higher resources (parallel \glspl{pe}) are allocated to boost performance. Before convolution, input is reorganized into kernel-size segments. Each \gls{pe} fetches weights/biases from on-chip memory and computes outputs using fixed-point arithmetic. The max-pooling layers and ReLU activations are optimized with \gls{simd} parallelism. For each input convolution feature map $(C_{\text{in}})$, $N_{\text{SIMD}}$ channels are processed per clock cycle using a dedicated activation unit that clamps negative values to zero. Each \gls{simd} lane concurrently processes one activation element from the segment and new input values are streamed in via the line-buffers. The folding factor for the $i^\text{th}$ ReLU layer is computed as $F^{(i)} = C_{\text{in}}^{(i)} \, / \, N_{\text{SIMD}}^{(i)}$. The batch-norm layer is merged with the preceding convolution layer. This approach reduces the utilization of \gls{bram} by eliminating the need to store layer parameters separately in on-chip memory. In addition, it minimizes the hardware resources required for the synthesis of the layer. This fusion is performed after the quantization-aware training, and the fused network parameters are exported for deployment on \gls{fpga}. 

\section{Evaluation Setup}
We selected two widely-used 3D object classification benchmarks, ModelNet40~\cite{modelnet40} and ScanObjectNN~\cite{scanobjectnn} for evaluation. 
For hardware deployment, we use PointMLP-Elite ~\cite{POINTMLP}, a state-of-the-art model that achieves an overall classification accuracy of 93.6\% on the ModelNet40 dataset. PointMLP-Elite captures local geometric structures in point cloud data using a local grouper module, and applies a learnable affine transformation to normalize them to a stable representation before local and global feature extraction. Topologically, the model is structured into four distinct stages of varying depths with 24 1D-convolution layers, a classifier head, and a geometric normalization module with two geometric parameters, \( \alpha \) and \( \beta \). The value of these parameters are determined during training alongside weights and activations, requiring extra resources for storage and computation.   

The PointMLP-Lite model experiments were carried out using Python version $3.10.14$, PyTorch version $2.4.0$, and CUDA version $12.4$ on an NVIDIA GeForce RTX 3090 \gls{gpu}. For quantization-aware training, Brevitas framework $0.7.0$ was used. We use \textit{stochastic gradient descent} optimizer with $momentum=0.8$ and $weightDecay = 0.0002$, and \textit{CosineAnnealingLR} scheduler with initial and minimum \textit{learning rate} values of $0.1$ and $0.005$ respectively. We train our models for $1,000$ epochs with a batch size of $256$. Vivado \gls{hls} 2018.3 was used for synthesis and \gls{fpga} deployment pipeline. The complete system was implemented using Vivado Block Design, targeting the Xilinx Zynq 7000 SoC ZC706 board. For \gls{fpga} power measurement, a socket power meter Voltacraft VC-870 was used. 

\section{Evaluation Results}

\begin{table}[t]
\centering
\caption{ PointMLP-Elite accuracy results on ModelNet40 \& ScanObjectNN.}
\label{tab:ablation_study}
\resizebox{9.4cm}{!}{
\begin{tabular}{cccccccccc}
\hline
\textbf{Model}  & \textbf{Num of} & \textbf{Geometric} & \textbf{Sampling} & \textbf{BN} & \multicolumn{2}{c}{\textbf{ModelNet40}} & \multicolumn{2}{c}{\textbf{ScanObjectNN}} \\
\textbf{}  & \textbf{Input} & \textbf{Param.} & \textbf{Algo.} & \textbf{Layer} &   &  &  & \\
\textbf{} & \textbf{Points} & \textbf{$\alpha$ \& $\beta$} & \textbf{} & \textbf{Fusion} & \textbf{OA(\%)}  & \textbf{mA(\%)} & \textbf{OA(\%)}  & \textbf{mA(\%)} \\
\hline
PointMLP-Elite~\cite{POINTMLP} & 1024 & \cmark & \gls{fps} & \xmark & 93.60 & 90.90 & 83.50 & 81.10 \\ 
$M-1$ & 1024 & \xmark & \gls{urs} & \cmark & 92.30 & 89.80 & 80.88 & 78.73 \\ 
$M-2$ & 512 & \xmark & \gls{urs} & \cmark & 91.69 & 88.96 & 80.22 & 77.94 \\ 
$M-3$ & 256 & \xmark & \gls{urs} & \cmark & 90.56 & 87.72 & 72.31 & 68.64 \\ 
$M-4$ & 128 & \xmark & \gls{urs} & \cmark & 89.59 & 86.87 & 68.91 & 65.38\\
\hline
\end{tabular}}
\end{table}

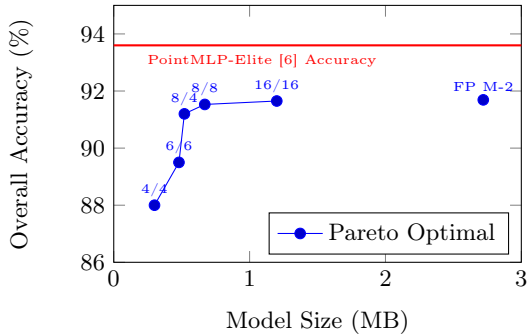
\begin{figure}[t]
\centering
\pgfplotsset{width=7cm,height=5cm}
\begin{tikzpicture}
\begin{axis}[ nodes near coords, xmin=0, xmax=3,  ymin=86, ymax=95, 
xlabel={Model Size (MB)}, ylabel={Overall Accuracy (\%)},legend pos=south east,
];
\addplot+[
point meta=explicit symbolic,
]
coordinates {

(0.3,88)[\tiny{4/4}] (0.48,89.5)[\tiny{6/6}] (0.52,91.2)[\tiny{8/4}] (0.67,91.53)[\tiny{8/8}] (1.2,91.65)[\tiny{16/16}]

(2.72,91.69)[\tiny{FP M-2}]
};

\draw [help lines,red,thick] (axis cs:0,93.6) -- (axis cs:3,93.6);
\node[red,below left] at (axis cs:2,93.6){\tiny{PointMLP-Elite~\cite{POINTMLP} Accuracy}};

\legend{Pareto Optimal};
\end{axis}
\end{tikzpicture}
\caption{Pareto frontier illustrating OA vs. model size trade-off on the ModelNet40.} \label{fig:pareto}
\end{figure}

Table~\ref{tab:ablation_study} summarizes the impact of various compression strategies on \gls{oa} and \gls{ma} for the ModelNet40 and ScanObjectNN datasets. The \gls{oa} metric shows the average
accuracy across all test instances, while \gls{ma} presents the mean accuracy across all shape classes. Starting from the original model with $N_{input} = 1024$, we experimented with reduced input sizes $N_{input} \in [512, 256, 128]$, pruning the 
\( \alpha \) and \( \beta \) parameters, replacing \gls{fps} with \gls{urs}, and fusing batch-norm with convolution layers. The models are labeled \textit{M-1} to \textit{M-4}. Notably, model \textit{M-2}, with 512 input points, shows only a $\sim$2\% drop in \gls{oa} on ModelNet40 and $\sim$3\% on ScanObjectNN, while halving the size of intermediate features. This significantly reduces the memory footprint, making it a practical trade-off for \gls{fpga} deployment. Based on these results, we select \textit{M-2} as the baseline for quantization. The Pareto-frontier chart in Fig.\ref{fig:pareto} illustrates the \gls{oa} vs. model size trade-off using the ModelNet40 dataset. Each data point in the figure is represented with
a \textit{weight-precision/activation-precision}. It can be seen that 8/8-bit quantized model is Pareto optimal with similar accuracy as compared to the \textit{M-2} model but with $4\times$ less complexity. Based on these findings, we introduce a new model called PointMLP-Lite. 

\begin{table}[t]
\caption{Comparison with previous 3D point cloud \gls{fpga} architectures. }
\begin{center}
\resizebox{!}{3.7cm}{
\begin{tabular}{l|l|l|l|l|l}
\toprule
    & SOCC 2022  & ISCAS 2020  & CSSP 2023  & ASICON 2019 &  HLS4PC   \\
    &~\cite{SSCN_FPGA} & ~\cite{POINTNET_FPGA} & ~\cite{DGCNN_FPGA} &~\cite{OPOINTNET_FPGA} &   (This Work)   \\
   \midrule

\textbf{Benchmarks}   & ShapeNet/  & -- & ModelNet40/ & -- & ModelNet40/ \\
   & NYU Depth  &  & ShapeNet2Core &  & ScanObjectNN  \\
   \midrule
\textbf{Topology}   & SSCN  & PointNet & DGCNN & O-PointNet & PointMLP-Lite \\ 
\midrule
\textbf{Conv layers} &  --  & 6 & 4 EdgeConv & 7 & 24 \\
\midrule
\textbf{MLP layers}  &  --  & 6 & 3 & 1 & 3 \\
\midrule
\textbf{Platform}    &  ZCU102  & ZCU104 &  Ultrascale V9UP & ZC706 & ZC706 \\
\midrule
\textbf{Architecture type}    &  Compute Array  & Process Element 
 & Systolic Array & Parallel &  Streaming-based     \\
    &    & (PE) Array  &  & Computing Unit & Dataflow    \\
\midrule
    
\textbf{Precision}   &  Int8  & Int8/Int16 & \gls{fp}32 & fp16 & fp8 \\
\midrule
\textbf{FF}          &  12.1K (2.22\%) & 36K (8\%)/60K (13\%) & 44.48\% & -- & 34k (8\%) \\ \midrule
\textbf{LUT}         &  17.6K (6.43\%) & 19K (8\%)/30K (13\%)&  78.92\% & -- & 92k (42\%)\\ \midrule
\textbf{DSP}         &  256 (10.16\%) & 1K (60\%)/1K (60\%)& 27.42\% & -- & 0 (0\%)\\ \midrule    
\textbf{BRAM}        &  365 (40.08\%) & 114 (37\%)/123 (39\%)& 39.2\% & -- & 401 (73\%)\\ \midrule
\textbf{URAM}        &  0 &  48 (50\%)/96 (100\%) & 0 & 0 & 0 \\ \midrule
\textbf{Frequency [MHz]}   & \textbf{270} & 100 & 130 & 100 & 100 \\ \midrule  
\textbf{Power Consumption [W]}   & 3.45 & -- & 17 & \textbf{2.14} & 2.2 \\ \midrule  
\textbf{Throughput [GOPS]}   & 17.73 & 182.1/130 & -- & 1.208 & \textbf{648} \\
\midrule
\textbf{Energy Efficiency [GOPS/W]}   & 5.13 & -- & -- & 0.56 & \textbf{294.5} \\ 
\bottomrule
\end{tabular}}
\label{tab:sota}
\end{center}
\end{table}

Table \ref{tab:sota} shows the \gls{fpga} deployment results of the PointMLP-Lite model using the HLS4PC framework and compares them with prior \gls{fpga}-based 3D point cloud accelerators in terms of resource utilization, throughput, power consumption, and energy efficiency. Note that previous accelerators differ in architecture, target models, and model complexity. Our framework achieves significantly higher throughput ($3.56\times$) and energy efficiency ($57.4\times$). These gains result from hardware-aware optimizations of the PointMLP-Lite model, and careful algorithm-hardware co-design while considering the compute and memory constraints of \glspl{fpga}. In contrast, prior works focused on deploying models directly with minimal optimizations (e.g. FP32 implementation of ~\cite{DGCNN_FPGA}). Furthermore, the reconfigurable layer-level parallelism of our library results in increased throughput and enables energy-efficient deployment even if it may affect accuracy to some extent.

Table \ref{tab:throughput_original} presents a comparison of throughput of \gls{cpu}, \gls{fpga}, and \gls{gpu} of our PointMLP-Lite model against \gls{gpu} implementation of the baseline PointMLP-Elite Model. 
Our proposed architecture achieves $5.6\times$ higher throughput as compared to the baseline \gls{gpu} implementation.

\begin{table}[ht]
\centering
\caption{Comparison of the \gls{fpga} architecture with baseline, \gls{cpu} and \gls{gpu}} \label{tab:throughput_original}
\resizebox{8.5cm}{!}{
\begin{tabular}{lllllllll} 
\toprule

Model & Platform & Frequency(GHz) & Throughput(SPS) \\
\midrule 
PointMLP-Elite (Baseline)~\cite{POINTMLP} & Tesla V-100 & 1.2 & 176 \\
PointMLP-Elite & RTX 3060 Ti & 2.1 & 187  \\
PointMLP-Lite (This Work) & RTX 3060 Ti & 2.1 & 421  \\
PointMLP-Lite (This Work) & Intel i5-13400 & 4.6 & 45  \\
PointMLP-Lite (This Work) & Xilinx ZC706 & 0.1 & \textbf{990}  \\
\bottomrule
\end{tabular}}
\end{table}

\section{Conclusion}\label{conclusion}
In this paper, we present HLS4PC, a parameterizable \gls{fpga}-based framework for accelerating point-based 3D point cloud models. We also propose a compressed PointMLP-Lite model, which uses 8-bit precision for both weights/activations and 512 input 3D points. The model integrates batch-norm into convolution layers, replaces \gls{fps} with \gls{urs}, and prunes geometric normalization parameters. PointMLP-Lite reduces complexity by \(4\times\) compared to the baseline PointMLP-Elite, while achieving a \(3.56\times\) higher throughput when deployed on an \gls{fpga} compared to prior works. As future work, we plan to explore Hilbert Curve-based sampling to reduce accuracy loss from \gls{urs}.

\bibliographystyle{splncs04} 
{\footnotesize
\bibliography{references}}

\end{document}